\documentstyle[12pt] {article}
\renewcommand{\thefootnote}{\fnsymbol{footnote}}
\def\zid{1\kern-0.36em\llap~1}

\newcommand{\beq}{\begin{equation}}
\newcommand{\ber}{\begin{eqnarray}}
\newcommand{\eeq}{\end{equation}}
\newcommand{\eer}{\end{eqnarray}}

\begin{document} \begin{titlepage} \rightline{\vbox{\halign{&#\hfill\cr
&\normalsize CERN-TH.6950/93\cr
&\normalsize ANL-HEP-PR-93-57\cr
&\normalsize August 1993\cr}}}
\vspace{.5in}
\begin{center}

\LARGE {\bf Transverse Momentum Distributions for
Heavy Quark Pairs 
}
\vskip   .6in
\large Edmond L. Berger$^{a,b}$ and Ruibin Meng$^a$  \\
\vskip .4in
$^a$High Energy Physics Division
\\Argonne National Laboratory, Argonne, IL 60439, USA\\
$^b$CERN, Geneva, Switzerland\\
\end{center}

\begin{abstract}
We study the transverse momentum distribution for a $pair$ of heavy quarks
produced in hadron-hadron interactions.  Predictions for
the large transverse momentum region are based on exact order $\alpha_s^3$
QCD perturbation theory.  For the small transverse momentum region, we
use techniques for all orders resummation of leading logarithmic contributions
associated with initial state soft gluon radiation.  The combination
provides the transverse momentum distribution of
heavy quark pairs for all transverse momenta.  Explicit results are presented
for $b\bar b$ pair production at the Fermilab Tevatron collider and for
$c\bar c$ pair production at fixed target energies.
\end{abstract}
\renewcommand{\thefootnote}{\arabic{footnote}} \end{titlepage}

\section {Introduction}  \label{1}

The distribution in the transverse momentum $q_\perp$ of a heavy quark $pair$,
$Q\bar Q$, produced in hadron-hadron interactions is of interest for
elucidating the underlying quantum
chromodynamics (QCD), and its understanding is important in
studies of $B\!\!-\!\!\bar B$ mixing and $C\!P$ violation at hadronic
facilities.  Unlike the case of pairs of heavy quarks produced in $e^+e^-$
annihilation, $Q\bar Q$ pairs created
in hadron hadron collisions are often not in a back-to-back configuration
(even in a plane transverse to the beam direction).  The net transverse
momentum
of the pair measures the imbalance between the transverse momenta of the
$\bar Q$ and the $Q$.  In this paper, we examine the quantitative description
in QCD perturbation theory of the expected imbalance.  Predictions for the
region of large $q_\perp$ are based on exact order $\alpha_s^3$ QCD
perturbation theory.
For the regions of small and modest $q_\perp$, we employ an all orders
resummation of leading logarithmic contributions associated with the emission
of soft gluons from the intial-state partons that participate in the
hard scattering process.  This calculation addresses a practical question
for heavy quark tagging at hadron facilities: if a $Q$ is tagged with a given
transverse momentum, what distribution in transverse momentum should one
expect to observe for the associated $\bar Q$ ?

We consider the process $hadron + hadron \rightarrow Q + \bar Q +X$.
In the simplest parton model description, the underlying hard scattering
process is $parton + parton \rightarrow Q + \bar Q$.  At this level of
approximation, no bremsstrahlung gluons are radiated from initial-state
or
final-state partons.  If one neglects intrinsic transverse momentum of
initial state partons, $q_\perp$ is zero.  Therefore, in the simplest
parton
model description, $Q$ and $\bar Q$  pairs would be
produced back-to-back in the transverse plane.  However, gluon
bremsstrahlung
is important in QCD and in general generates non-zero $q_\perp$ .

The single particle inclusive differential cross section for heavy
quark production has been studied in detail at next-to-leading order
in QCD.  We cite in particular the calculation of the
complete first order corrections to the dominant QCD production
channels \cite{nde,admn} and \cite{bkns,bnmss,rm} and
comparisons with data on inclusive b-quark production
from the UA1\cite{UA1} and CDF collaborations\cite{CDF}.
We mention also the recent work on resummation of
leading logarithmic contributions for the one particle inclusive
cross section\cite{LSN1}.  In the single particle
inclusive approach, the kinematical variables of the
heavy quark's (or antiquark's) partner and of the final state
light partons are integrated over with the attendant
limitation that it is not possible to examine quark-antiquark
correlations or the cross section differential in the transverse
momentum $q_T$ of the $Q\bar Q$ pair.  Correlations have been studied at
leading
order\cite{elb1} and at next-to-leading order\cite{kuebel,mnr}.
One may expect that next-to-leading order QCD should provide reliable
expectations for the distribution in $q_T$ at large $q_T$.  At small $q_T$,
the relatively large mass $m_Q$ of the heavy quark $Q$ justifies perturbation
theory, but the presence of the two disparate scales, $m_Q$ and $q_T$, requires
care.

Perturbative QCD has been used for a successful description of transverse
momentum distributions in massive lepton-pair production, the Drell-Yan
process\cite{dy}.  There one studies the reaction $hadron+hadron \rightarrow
e^+ + e^- + X$, where the electron-positron pair is detected and
its transverse momentum $q_\perp$ is measured\cite{cs1,dsw,na10,e615}.
Important for the quantitative description of the
transverse momentum distribution at modest values of $q_\perp$ is the
resummation of logarithmic contributions
associated with emission of soft gluons in the
initial state of the hard scattering process $quark+antiquark \rightarrow
e^+ + e^- + X$.  We follow closely the analogy with the Drell-Yan case
for the reaction $hadron+hadron \rightarrow Q + \bar Q + X$ and concentrate on
the transverse momentum distribution of the $Q\bar Q$ quark
pair.  Heavy quark pair production is, however, more
involved than massive lepton-pair production.
New complications arise from soft gluon emission from the
final-state heavy quarks, effects that are absent in the Drell-Yan reaction.
We will argue that the resummation technique for dealing with initial gluon
radiation should still be applicable in our case.  Soft gluon emission from
final state heavy quarks has been studied in \cite{MW}.

For the process $hadron + hadron \rightarrow Q + \bar Q +X$, we
choose to study the cross
section differential in the variables $M^2$ and $q_\perp^2$;
$M$ is the invariant mass of the $Q\bar Q$  pair.
Knowing $M$ and $q_\perp$ ,
we can judge how far the $Q\bar Q$ system is away from
the back-to-back configuration in the transverse plane.
To be more specific, when $M$ is near the
mass threshold $2m$ of the $Q \bar Q$ pair, the momenta of the $Q$
and $\bar Q$ are close to zero in the center of mass frame of the $Q\bar Q$
system.  Only a small amount of $q_\perp$ then suffices to put the
$Q$ and $\bar Q$ in a configuration that is non-back-to-back in the
transverse plane in the laboratory system of reference.
On the other hand, at large $M$, the $Q$ and
$\bar Q$ have large relative momentum in the center of mass
frame of the pair.  At large $M$, large $q_\perp$ would be
needed to produce a $Q \bar Q$ pair that is not-back-to-back in the
transverse plane in the laboratory system.

At large and moderate values of $q_\perp^2 \sim {\cal O}(M^2)$,
the $Q \bar Q$ pair production cross section
can be computed perturbatively as
\begin{eqnarray}
{d\sigma \over dM^2 dq_\perp^2}
=\alpha_s^3(a_1+a_2\alpha_s+a_3\alpha_s^2+\cdots).
\label{EQ:PERTI}
\end{eqnarray}
At any fixed  order of $\alpha_s$ and  $q_\perp^2\ne 0$,
the cross section is well behaved after the hard scattering
cross section has been properly defined.  At low $q_\perp^2\ne 0$, however,
the convergence of the perturbative series
deteriorates.  For small $q_\perp^2 \ne 0$, the dominant contributions
(i.e. the leading
logarithmic contributions) to Eq.~(\ref{EQ:PERTI}) have the form
\begin{eqnarray}
\frac{d\sigma}{dM^2dq_\perp^2} \sim
 \frac{\alpha_s^2}{q_\perp^2}\left(b_1\alpha_s\ln(\frac{M^2}{q_\perp^2})
+ b_2\alpha_s^2\ln^3(\frac{M^2}{q_\perp^2})+\cdots\right).
\label{EQ:LLC}
\end{eqnarray}
The convergence of the theory is therefore
governed by $\alpha_s\ln^2(M^2/q_\perp^2)$
instead of $\alpha_s$.
The logarithms arise through emission of
soft and collinear gluons.
At sufficiently low $q_\perp^2$, $\alpha_s\ln^2(M^2/q_\perp^2)$
is large even when $\alpha_s$ is small and
any fixed order calculation
breaks down.  In order to obtain a reliable prediction, one
must resum the leading contributions
to all orders in $\alpha_s$.

The remainder of this paper is organized as follows. In Sec.~2 we
present the perturbative calculation of the $q_\perp^2$
distribution using exact order $\alpha_s^3$ QCD matrix elements.
We describe how to obtain the asymptotic expression
at $q_\perp^2\rightarrow 0$ from the exact order $\alpha_s^3$ matrix elements.
In Sec.~3 we discuss the formalism for resumming the initial soft and
collinear gluon contributions.
We match  the resummed result in the low-$q_\perp^2$ region to
the exact ${\cal O}(\alpha_s^3)$ result in the high-$q_\perp^2$
region.  Results and examples of the $q_\perp^2$
distribution for specific hadronic reactions are given in Sec.~4.

\section{Perturbative Calculation}

We begin our discussion with the hadronic reaction in
which a $Q\bar Q$ pair is produced:
\begin{eqnarray}
p(K_1) + \bar p(K_2) \rightarrow Q(p_1) + \bar Q(p_2) + X
\label{EQ:HADRON}
\end{eqnarray}
In this expression, $p$ and $\bar p$ denote the proton and antiproton
respectively.  The quantity $X$ stands for all the final particle states that
we sum over so that the above process is semi-inclusive with respect to the
outgoing final particles.  We use capital letters for the momenta of the
proton and antiproton to distinguish them from those of the quarks,
antiquarks and gluons.  The corresponding partonic subprocess can be
written as
\begin{eqnarray}
p_i(k_1) + p_j(k_2) \rightarrow Q(p_1) + \bar Q(p_2) + X,
\end{eqnarray}
\label{EQ:PARTON}
where $p_{i,j}$ are the initial partons from the proton and
the antiproton.
The relationship between the momenta of the hadrons and partons is
\begin{eqnarray}
k_1=\xi_1K_1,\> \quad k_2=\xi_2K_2
\label{EQ:MF}
\end{eqnarray}
The four-momentum of the system made up of the $Q\bar Q$ pair is
\begin{eqnarray}
q^\mu =p_1^\mu+p_2^\mu
\end{eqnarray}
\label{EQ:MPAIR}
We also define several useful invariants for our calculation:
\begin{eqnarray}
S&=&(K_1+K_2)^2  \nonumber \\
s&=& (k_1+k_2)^2 \;=\;\xi_1 \xi_2 S  \nonumber \\
M^2\!&=&q^2\;=\;(p_1+p_2)^2 \nonumber \\
x_1\!&=&\frac{M^2}{2q\cdot K_1},\>\quad  x_2\;=\;\frac{M^2}{2q\cdot K_2} \\
Y  &=&\frac{1}{2}\ln({q\cdot K_2}/{q\cdot K_1})
\;\equiv \;y + \frac{1}{2}\ln(\xi_1/\xi_2)  \nonumber \\
q_\perp^2&=&\frac{2q\cdot K_1\,q\cdot K_2}{K_1\cdot K_2}-M^2
\;=\;\frac{2q\cdot k_1\,q\cdot k_2}{k_1\cdot k_2}-M^2 \nonumber
\label{EQ:INVARIANTS}
\end{eqnarray}
We remark that
$\sqrt{S}$ is the collision energy of the proton and antiproton;
$M$ is the invariant mass of the heavy quark $Q\bar Q$ pair;
$q_\perp^2$ is the square of the transverse momentum of the pair,
equal to the square of the vector sum of the
individual transverse momenta of the $Q$ and $\bar Q$;
$Y$ is the rapidity of the $Q\bar Q$ pair in the
center of mass frame of the proton and antiproton; and
$y$ is the rapidity of the $Q\bar Q$ pair in the
center of mass frame of the interacting initial partons.

The differential cross section for reaction (\ref{EQ:HADRON}),
is expressed as a convolution of partonic cross sections and parton
distribution functions:
\begin{eqnarray}
\frac{d^2\sigma_{p\bar p\rightarrow Q\bar QX}}{dM^2dq_\perp^2 dY}
= \sum_{i,j=q,\bar q,g}
\int^1_{\tau_{min}}\!\!{d\xi_1} \int^1_{\tau_{min}/\xi_1}\!\!\!{d\xi_2}
\;\, f_{i/p}(\xi_1,\mu)\,f_{j/\bar p}(\xi_2,\mu)
\times \frac{d^2\hat\sigma_{ij\rightarrow Q\bar QX}}{dM^2dq_\perp^2 dy}.
\label{EQ:CONV}
\end{eqnarray}
Here $\tau_{min}=(\sqrt{M^2+q_\perp^2}+q_\perp)^2/S$, and
${d^2\hat\sigma_{ij\rightarrow Q\bar QX}}/dM^2dq_\perp^2dy$
is the fixed-order reduced partonic cross section
obtained by first calculating Feynman diagrams up to a given
fixed-order in QCD perturbation theory and then implementing a renormalization
scheme to remove any ultraviolet divergences. The soft
divergences at $q_\perp\rightarrow 0$
are cancelled between the virtual diagrams and the
bremsstrahlung diagrams. The collinear divergences (mass singularities)
at $q_\perp\rightarrow 0$
are absorbed into the definition of the parton distribution functions.
Therefore the fixed-order reduced partonic cross section is well behaved
for any value of $q_\perp^2$, in particular $q_\perp^2=0$.
(Note that the apparent divergence at $q_\perp^2=0$ in Eq. (2) is
cancelled by a $\delta$-function that arises from virtual
diagrams.)
An advantage of calculating
the order $\alpha_s^3$ differential
cross section in terms of $q_\perp^2$
is that $q_\perp^2$ can be used
as a cutoff variable for both the infrared and collinear
divergences in the bremsstrahlung diagram calculations.
This means that at any finite value of $q_\perp^2$, the
order $\alpha_s^3$ differential cross section
can be calculated by evaluating
the bremsstrahlung diagrams without
explicit concern about soft gluon cancellation
and factorization of the collinear singularity.

We now consider $Q\bar Q$ pair production in QCD
perturbation theory.  It proceeds by the following two reactions
in the Born approximation (order $\alpha_s^2$):
\begin{eqnarray}
q(k_1) + \bar q(k_2) \rightarrow Q(p_1) + \bar Q(p_2) \nonumber\\
g(k_1) + g(k_2)      \rightarrow Q(p_1) +  \bar Q(p_2)
\label{EQ:BORNP}
\end{eqnarray}
Symbols inside the parentheses denote the momentum assignments for the
partons. The Feynman graphs that contribute to the Born amplitude are
shown in Fig. 1.

The magnitude squared of the Born amplitude, averaged over initial
colors and spins and summed over final colors and spins, can be
written as\cite{gor}
\begin{eqnarray}
\overline{|M|}^2_{q\bar q\rightarrow Q\bar Q}&=&
\frac{g^4_s}{N}C_F A_{QED} \nonumber \\
\overline{|M|}^2_{gg\rightarrow Q\bar Q}&=&
\frac{g^4_s}{N^2-1} \left(  C_F - C_A\frac{t_1u_1}{s} \right) B_{QED}
\label{EQ:BAM}
\end{eqnarray}
with
\begin{eqnarray}
A_{QED} &=& \frac{t_1^2+u_1^2}{s}  +\frac{2m^2}{s} \nonumber\\
B_{QED} &=& \frac{t_1}{u_1} + \frac{u_1}{t_1} + \frac{4m^2 s}{t_1u_1}
      (1-\frac{m^2 s}{t_1u_1}); \nonumber \\
\label{EQ:AB}
\end{eqnarray}
$t_1$ and $u_1$ are defined as
\begin{eqnarray}
t_1= (k_2-p_2)^2-m^2, \>\quad u_1= (k_1-p_2)^2-m^2.
\label{EQ:T1U1}
\end{eqnarray}
It is easy to verify that
\begin{eqnarray}
\frac{d\sigma^{(0)}_{q\bar q,gg\rightarrow Q\bar Q}}
{dM^2dq_\perp^2dy} &=&
\delta(1-\frac{x_1}{\xi_1})\,\delta(1-\frac{x_2}{\xi_2})\, \delta(q^2_\perp)\,
\frac{\sigma^{(0)}_{q\bar q,gg}(M^2)}{M^2}.
\label{EQ:BCS}
\end{eqnarray}
The total partonic cross sections are\cite{ndet}
\begin{eqnarray}
\sigma^{(0)}_{q\bar q}(M^2) &=&\frac{2\pi\alpha_s^2}{3 M^2}\frac{C_F}{N}
\beta \,(1+2m^2/M^2) \\
\sigma^{(0)}_{gg}(M^2) &=&\frac{2\pi\alpha_s^2}{ M^2}\frac{C_F}{N^2-1}
\left[ -(1+\frac{4m^2}{M^2})\beta
-(1+4\frac{m^2}{M^2}-\frac{8m^4}{M^4})
\ln x \right] \nonumber\\
&&+\frac{2\pi\alpha_s^2}{M^2}\frac{C_A}{N^2-1}
\left[ (-\frac{1}{3}-\frac{5}{3}\frac{m^2}{M^2}
)\beta - \frac{4m^4}{M^4} \ln x \right],
\label{EQ:BTCS}
\end{eqnarray}
with
\begin{eqnarray}
\beta = \sqrt{1-4m^2/M^2}, \>\quad x=\frac{1-\beta}{1+\beta}.
\label{EQ:XBETA}
\end{eqnarray}
The $C_A$ and $C_F$ are the Casimir invariants for
the adjoint and the fundamental representation of $SU(N)$.  For the
particular case of $SU(3)$,
\begin{eqnarray}
N=3,\>\quad C_A=N,\>\quad C_F=\frac{N^2-1}{2N}.
\label{EQ:CACF}
\end{eqnarray}
Observe that the differential cross section Eq.~(\ref{EQ:BCS}) is
proportional to $\delta(q_\perp^2)$, because in the Born approximation
the heavy quark pairs are produced with $q_\perp^2=0$.
The total partonic cross sections $\sigma^{(0)}_{q\bar q}(M^2)$
and  $\sigma^{(0)}_{gg}(M^2)$ are inversely proportional to $M^2$.
This dependence can be understood because the lowest order
graphs either have only s-channel poles or fermion exchange
lines.  Most of the heavy quark pairs are therefore expected to be produced
near threshold, $M^2=4m^2$.

At order $\alpha_s^3$ the contributing partonic subprocesses
include gluon bremsstrahlung diagrams and (anti)quark-gluon scattering
diagrams:
\begin{eqnarray}
q(k_1)+\bar q(k_2) & \rightarrow & g(k_3) + Q(p_1) + \bar Q(p_2)
\nonumber \\
g(k_1)+g(k_2) & \rightarrow & g(k_3) + Q(p_1) + \bar Q(p_2) \nonumber\\
g(k_1)+q(\bar q)(k_2) & \rightarrow & q(\bar q)(k_3) + Q(p_1) + \bar Q(p_2)
\nonumber \\
\label{EQ:ALFSP}
\end{eqnarray}
plus virtual correction diagrams\cite{virtual}
to the lowest order processes Eq.(\ref{EQ:BORNP}).
In Fig. 2 we show some examples of the gluon bremsstrahlung
diagrams that contribute at order $\alpha_s^3$.

In the order $\alpha_s^3$ processes (\ref{EQ:ALFSP}),
$q_\perp^2$ is no longer constrained to be zero;
a spectrum of values of $q_\perp^2$ will be produced.
The $q_\perp^2\rightarrow 0$ limit means that
the parton with momentum $k_3$ in the processes (\ref{EQ:ALFSP}),
is either soft and/or collinear to
one of the initial or final state partons.
We note that in the $q\bar q$ channel there are diagrams with
a gluon emitted from an initial quark or antiquark line; these
look diagrammatically exactly the same as corresponding graphs
in the Drell-Yan reaction.  In addition there are diagrams
with a gluon emitted from a final heavy quark or antiquark line;
these are absent in the Drell-Yan reaction.
The initial gluon emission diagrams
can be soft and/or collinear
divergent at $q_\perp^2\rightarrow 0$.
The final gluon emission diagrams
can have only a soft divergence at $q_\perp^2\rightarrow 0$
because the final state quarks are massive.
The same statements can be made for the
gluon gluon initiated subprocesses which are usually the dominant
processes for $Q\bar Q$ pair production.
The quark-gluon scattering diagrams have only a collinear divergence
from light quark emission.
The square of the matrix element for the
order $\alpha_s^3$ processes Eq.~(\ref{EQ:ALFSP})
has been published\cite{es-gk}.
We will make use of those results in
our calculation.

To calculate the partonic differential cross section,
$d\sigma^{(1)}_{ij\rightarrow Q\bar Qk}/dM^2dq_\perp^2 dy$,
at order $\alpha_s^3$ we must integrate
over variables which are independent of $M^2$, $q^2_\perp$ and $y$
for the gluon bremsstrahlung and quark-gluon scattering processes.
We choose the $Q\bar Q$ center of mass frame in which
\begin{eqnarray}
q^{\mu}&=&(p_1+p_2)^{\mu}\;=\; (M,\; 0,\; 0,\; 0)  \nonumber \\
p_1^{\mu}&=&(M/2, \;\;    \omega_0\sin{\theta_1}\sin{\theta_2},
                  \;\;   \omega_0\sin{\theta_1}\cos{\theta_2},
                  \;\;   \omega_0\cos{\theta_1})\nonumber \\
p_2^{\mu}&=&(M/2,    - \omega_0\sin{\theta_1}\sin{\theta_2},
                     -\omega_0\sin{\theta_1}\cos{\theta_2},
                     -\omega_0\cos{\theta_1})  \nonumber \\
\label{EQ:CMQQB}
\end{eqnarray}
with
\begin{eqnarray}
\omega_0=\frac{M}{2}\sqrt{1-4m^2/M^2}.
\label{EQ:OMEGA}
\end{eqnarray}
We obtain
\begin{eqnarray}
\frac{d\sigma^{(1)}_{ij\rightarrow Q\bar Q k}}{dM^2dq_\perp^2dy}({\rm pert})
 &=& \frac{1}{128\,\pi^4\, s M^2}\, \frac{x_1}{\xi_1} \frac{x_2}{\xi_2}
\,\;\delta\left((1-\frac{x_1}{\xi_1})(1-\frac{x_2}{\xi_2})
- \frac{q_\perp^2/M^2}
{1+q_\perp^2/M^2}\right)  \nonumber\\
&&\times \int\frac{d^3p_1}{2p^0_1}\:
\delta\left((q-p_1)^2-m^2\right)\:\overline{|M|}^2_{ij\rightarrow Q\bar Qk}.
\label{EQ:ANL}
\end{eqnarray}
The integral $\int d^3p_1$ can be expressed explicitly as
\begin{eqnarray}
\int\!\frac{d^3p_1}{2p^0_1}\:
\delta\left((q-p_1)^2-m^2\right)\:\overline{|M|}^2_{ij\rightarrow Q\bar Qk}
= \frac{1}{8}\sqrt{1-4m^2/M^2}
 \int^\pi_0\!d\theta_1\sin\theta_1
 \int^{2\pi}_0\!d\theta_2\;\overline{|M|}^2_{ij\rightarrow Q\bar Qk},
\nonumber \\
\label{EQ:ANLINT}
\end{eqnarray}
and $\overline{|M|}^2_{ij\rightarrow Q\bar Qk}$ is square of the
matrix element for the order $\alpha_s^3$ processes (\ref{EQ:ALFSP}),
averaged over spin and color.

It is possible to perform the angular integrals analytically.
However, the squared matrix elements
for the $Q\bar Q$ production process are lengthy
and the integrals are difficult, making the calculation formidable.
Fortunately, for any finite value of $q_\perp^2$
the order $\alpha_s^3$ differential cross
section Eq.~(\ref{EQ:ANL}) is free from any soft and collinear divergences
so that the angular integrals Eq.~(\ref{EQ:ANLINT}) can
be dealt with numerically.
Then the differential cross section for the proton antiproton
reaction Eq.~(\ref{EQ:HADRON}) can be calculated by convoluting
the partonic cross section Eq.~(\ref{EQ:ANL})
with the parton distribution functions according to Eq.~(\ref{EQ:CONV}).

A quantity of interest and importance is
the average of the square of the transverse momentum at a fixed invariant
mass $M$.  It can be calculated as\cite{app}
\begin{eqnarray}
&&<q_\perp^2(M)>\left(\frac{d\sigma_{p\bar p\rightarrow Q\bar QX}}
{dM^2}\right) =
\!\int dq_\perp^2\,q_\perp^2\,\frac{d^2\sigma_{p\bar p
\rightarrow Q\bar QX} }
{dM^2dq_\perp^2} \nonumber \\
&&\qquad = \sum_{i,j=q,\bar q,g}
\int^1_{M^2/S} \!{d\xi_1} \int^1_{M^2/(\xi_1S)}\!\! {d\xi_2}
\,f_{i/p}(\xi_1,\mu)\,f_{j/\bar{p}}(\xi_2,\mu)
\int_0^{( s-M^2)^2/(4s)}
\!\!dq_\perp^2\,q_\perp^2\,\frac{d^2\sigma^{(1)}_{ij\rightarrow Q\bar Qk}}
{dM^2dq_\perp^2}, \nonumber \\
\label{EQ:AVGPM}
\end{eqnarray}
where
\begin{eqnarray}
\frac{d\sigma_{p\bar p\rightarrow Q\bar QX}}{dM^2} &=&
\sum_{i,j=q,\bar q,g}
\!\int^1_{M^2/S} \!{d\xi_1} \int^1_{M^2/(\xi_1S)}\!\! {d\xi_2}
\,\, f_{i/p}(\xi_1,\mu)\, f_{j/\bar p}(\xi_2,\mu)
\left[\frac{d\hat\sigma^{(0)}_{ij\rightarrow Q\bar Q}}{dM^2} +
\frac{d\hat\sigma^{(1)}_{ij\rightarrow Q\bar QX}}{dM^2}\right] \nonumber \\
&\equiv &{\cal K}
\sum_{i,j=q,\bar q,g}\!
\!\int^1_{M^2/S} \!{d\xi_1} \int^1_{M^2/(\xi_1S)}\!\! {d\xi_2}
\,\, f_{i/p}(\xi_1,\mu)\, f_{j/\bar p}(\xi_2,\mu)
\frac{d\hat\sigma^{(0)}_{ij\rightarrow Q\bar Q}}{dM^2}.
\label{EQ:PMDIS}
\end{eqnarray}

Symbol ${\cal K}$ denotes the familiar K-factor,
not necessarily a constant in this case.
The transverse momentum averaged over all values of
invariant mass can also be calculated
in the same way.

As remarked in the Introduction, the convergence of the perturbation
series deteriorates in the region $q_\perp^2\ll M^2$.  For
predictions of improved reliability in that domain, one must try to
sum leading contributions from all orders in $\alpha_s$.  We will
use the procedure for resumming contributions from initial state soft
and collinear gluons developed by Collins and Soper.  To begin, we must
first extract the leading contributions in the region of small
$q_\perp^2\ll M^2$ from the $\alpha_s^3$ cross section.  This is done
by calculating the angular integrals in Eq.~(\ref{EQ:ANLINT})
analytically.  Fortunately, in the limit $q_\perp^2/M^2 \rightarrow 0$
the squared matrix element simplifies substantially
and the angular integrals can be calculated analytically
without too much difficulty\cite{bkns}.

We take the limit $q_\perp^2\rightarrow 0$
of the exact expressions for the square of the order $\alpha_s^3$
matrix elements and do the integral under the same limit.
The calculation is accomplished in two steps.
First, we calculate the soft gluon contributions by
setting the gluon momentum $k_3\rightarrow 0$ everywhere
in the matrix elements except in the denominators that
are singular as $k_3 \rightarrow 0$.
The soft gluon matrix elements have
also been derived in the literature
(see Eqns.(5.1-5) in Ref.\cite{bkns}
and  Eqns.(2.24-26) in Ref.\cite{bnmss}).
In the limit $k_3\rightarrow 0$,
the $2 \rightarrow 3$ kinematics can also be approximated
by $2 \rightarrow 2$ kinematics, which we can implement effectively  by
replacing the $\delta -$function in Eq.~(\ref{EQ:ANL}) by
\begin{eqnarray}
\delta\left((1-\frac{x_1}{\xi_1})(1-\frac{x_2}{\xi_2}) - \frac{q_\perp^2/M^2}
{1+q_\perp^2/M^2}\right)
\>\Longrightarrow \>
\ln(\frac{M^2}{q_\perp^2})
\,\delta (1-\frac{x_1}{\xi_1})\,\delta(1-\frac{x_2}{\xi_2}).
\label{EQ:REPLSF}
\end{eqnarray}
The singularities in the denominator can be replaced by
\begin{eqnarray}
(1-\frac{x_1}{\xi_1})(1-\frac{x_2}{\xi_2})\>\Longrightarrow \>
\frac{q_\perp^2}{M^2}.
\end{eqnarray}
Second, the hard collinear contributions can be calculated if we
replace the first $\delta -$function in Eq.~(\ref{EQ:ANL}) by
\begin{eqnarray}
\delta\left((1-\frac{x_1}{\xi_1})(1-\frac{x_2}{\xi_2}) - \frac{q_\perp^2}
{1+q_\perp^2/Q^2}\right)
\>\Longrightarrow \>
\frac{\delta (1-x_1/\xi_1)}{(1-x_2/\xi_2)_+}
+\frac{\delta (1-x_2/\xi_2)}{(1-x_1/\xi_1)_+}
\label{EQ:REPLCL}
\end{eqnarray}
The ``$+$'' prescription in the above equation, defined as
\begin{eqnarray}
\int_0^1 \frac{f(x)}{(1-x)_+}dx  = \int_0^1 \frac{f(x)-f(1)}{1-x}dx,
\end{eqnarray}
ensures that there is no double counting
in the phase space region where the soft and
collinear divergences overlap.
Both replacements, Eq.~(\ref{EQ:REPLSF}) and
Eq.~(\ref{EQ:REPLCL}), imply significant simplification for our calculation.
Their origin has been well demonstrated, e.g. in
Eqns.(2.11-13) of Ref.~\cite{kauff}.
After we perform the angular integrals and convolution with
the parton distribution functions we obtain the asymptotic
expression for the differential cross section for process (\ref{EQ:HADRON})
in the form
\begin{eqnarray}
{d^3 \sigma_{p\bar p\rightarrow Q\bar QX}
  \over dM^2\, dq_\perp^2\, dY}({\rm asym})&=&
\sum_{i,j=q,\bar q,g}\,
\sigma^{(0)}_{ij} (M^2)\, {1\over S}\,
{\alpha_s(M^2) \over 2 \pi} \, {1\over q_\perp^2}
\Biggl\{ \Bigl[ {\cal A}_{ij}^{(1)} \ln({M^2\over q_\perp^2})
 + {\cal B}_{ij}^{(1)} \Bigr]
f_{i/p}(x_1) f_{j/\bar p}(x_2)  \nonumber \\
&&\quad  + \Bigl[f_{i/p}(x_1) \Bigl( P_{j\leftarrow b}
                   \otimes f_{b/\bar p} \Bigr) (x_2)
 		+ \Bigl( P_{i\leftarrow a}
                   \otimes f_{a/p} \Bigr) (x_1) f_{j/\bar p}(x_2)
                   \Bigr] \Biggr\} \>.  \nonumber \\
\label{EQ:ASYM}
\end{eqnarray}
The symbol $\otimes$ denotes a convolution of the parton distribution function
$f$ and Altarelli-Parisi splitting function $P$, defined by
\begin{eqnarray}
\Bigl( f \otimes P \Bigr) (x) \equiv \int_x^1
 f(y)\> P\Bigr({x \over y}\Bigl)\> {dy\over y} \>.
\label{EQ:CONVOLUTION}
\end{eqnarray}
The functions ${\cal A}_{ij}^{(1)}$ come purely from initial
state soft  and collinear gluon radiation:
\begin{eqnarray}
{\cal A}_{q\bar q}^{(1)} = 2 C_F, \> \quad
{\cal A}_{gg}^{(1)} = 2 C_A,  \> \quad
{\cal A}_{q(\bar q)g}^{(1)} = 0
\label{EQ:ACAL}
\end{eqnarray}
The functions ${\cal B}^{(1)}_{ij}$ can be split into two parts,
in terms of initial and final state gluon radiation:
\begin{eqnarray}
{\cal B}_{ij}^{(1)} = B_{ij,I}^{(1)} + B_{ij,F}^{(1)}
\label{EQ:BCAL}
\end{eqnarray}
with
\begin{eqnarray}
B_{q\bar q,I}^{(1)} &=& - 3 C_F, \>\quad
B_{q\bar q,F}^{(1)} \;=\; 2 C_F + 2 C_F
\frac{\ln x}{\beta} (1 -2m^2/M^2) \nonumber\\
B_{gg,I}^{(1)} &=& - 2 \beta_0, \>\quad
B_{gg,F}^{(1)} \;=\; 2 C_F + 2 C_F
\frac{\ln x}{\beta} (1 -2m^2/M^2) \nonumber \\
B_{q(\bar q)g,I}^{(1)}\!\! &=& 0, \>\quad
\quad B_{q(\bar q)g,F}^{(1)} \;=\; 0 \nonumber\\
\label{EQ:BIF}
\end{eqnarray}
Quantities $x$  and $\beta$ have been defined in Eq.~(\ref{EQ:XBETA}).
Our coefficient $\beta_0$ of the $\beta$-function is
normalized to
\begin{eqnarray}
\beta_0 = \frac{11N - 2N_f}{6}.
\label{EQ:BETA0}
\end{eqnarray}
To verify our calculation
we have checked that the $1/\epsilon^2$,  $1/\epsilon$  pole terms generated
by dimensional regularization from  the terms not proportional
to the Altarelli-Parisi splitting functions in the expression
Eq.~(\ref{EQ:ASYM}) cancel the infrared pole terms
from virtual diagram calculations\cite{bkns,bnmss}.

The expressions for the initial state gluon radiation terms
for the $q\bar q$ channel, ${\cal A}_{q\bar q}^{(1)}$ and
$B_{q\bar q,I}^{(1)}$, are exactly the same as those in the Drell-Yan
reaction.  For initial state gluon radiation in the
$gg$ channel, our expressions ${\cal A}_{gg}^{(1)}$ and  $B_{gg}^{(1)}$
also agree with those found by Catani, D'Emilio, and Trentadue\cite{catani1}
and by Kauffman\cite{kauff}
for production of a color singlet state from gluon-gluon fusion.
The agreement in both cases follows the expectation that the initial state
soft gluon radiation does not depend on the type of
hard process under consideration\cite{ddt,LSN}.
It indicates that the effects of initial state gluon radiation
can be resummed to all orders of $\alpha_s$ for $Q\bar Q$ pair production,
as in the cases of the Drell-Yan reaction\cite{cs1,dsw} or Higgs production
through gluon fusion\cite{hinch,catani1,kauff}.  In the next section we will
attempt the resummation of the initial soft and collinear gluon contributions
using the formalism developed for the Drell-Yan process\cite{pp,cs2}.

The expressions for final state soft gluon radiation,
$B_{q\bar q,F}^{(1)}$ and $B_{gg,F}^{(1)}$,
are the same for the $q\bar q$ and $gg$ channels.
This can be understood since the final state soft gluon radiation
occurs from the final state (anti)heavy quark lines in both
the $q\bar q$ and $gg$ channels.
Soft gluon emission from final state heavy quarks
has been studied in Ref.~\cite{MW}.

\section {Resummation}  \label{2}

The technique for resumming contributions from initial state soft and
collinear gluons was developed by Collins and Soper\cite{cs2},
and it has been applied
to massive lepton-pair production\cite{dsw}, single vector boson
production\cite{keith,ak}, Higgs boson production\cite{hinch,catani1,kauff},
and ZZ-pair production\cite{hmo}.  In our case,
the appropriate expression analogous to that of Collins and Soper is
\begin{eqnarray}
{d^3\sigma \over dM^2\, dq_\perp^2\, dY}\, (\hbox{\rm resum})
&=& \sum_{i,j=q,\bar q,g}\,
\sigma^{(0)}_{ij} (M^2)\> {1\over 2S}\>
 \int_0^\infty \, db\, b\, J_0(b\, q_\perp) \> W_{ij} (M,b) \>,
\label{EQ:RESUM}
\end{eqnarray}
where $J_0(x)$ is the zeroth order Bessel function.
The function $W_{ij}(M,b)$ sums all the logarithmic terms of the
form $\alpha_s^n ln^m(M^2b^2)$ with $1\leq m\leq 2n$ in the impact parameter
$b$
space.  The all orders structure of $W$ is given by the functional form
\begin{eqnarray}
W_{ij}(M,b) &=&
\exp \Biggl\{ - \int_{C_1^2/b^2}^{C_2^2 M^2}\, {dq^2 \over q^2}\,
 \Bigl[ \ln \Bigl( {C_2^2 M^2 \over q^2} \Bigr) \,
 A_{ij} \bigl( \alpha_s (q^2) \bigr)
 + B_{ij} \bigl( \alpha_s (q^2) \bigr) \Bigr] \Biggr\} \nonumber \\
& & \times
\Bigl(C\otimes f_{i/p}\Bigr)(x_1,{C_3^2\over b^2}) \,\,\,
\Bigl(C\otimes f_{j/\bar p}\Bigr)(x_2,{C_3^2\over b^2})
\label{EQ:WEXP}
\end{eqnarray}
The parameters $C_1$, $C_2$ and $C_3$ are somewhat arbitrary.  They are
associated with the choices of renormalization and factorization scales
in a fixed order perturbative calculation.  We use the standard choices
\begin{eqnarray}
C_1=C_3=2e^{-\gamma_E}=b_0,\> \quad C_2=1;
\label{EQ:C123}
\end{eqnarray}
where $\gamma_E$ is Euler's constant.
The symbol $\otimes$ denotes a convolution, defined in Eq.~(30).
In the limit $q_\perp^{\phantom{7}}/M \rightarrow 0$, the
parton momentum fractions are
\begin{eqnarray}
x_1 = e^{ Y} \sqrt{ {M^2\over S} } \>, \quad
x_2 = e^{-Y} \sqrt{ {M^2\over S} } \>.
\label{EQ:X1X2}
\end{eqnarray}
The functions $A$, $B$, and $C(x)$ may be expanded in a perturbation series
in $\alpha_s$:
\begin{eqnarray}
A_{ij}(\alpha_s)&=&\sum^\infty_{n=1}A_{ij}^{(n)}\left(\frac{\alpha_s}{2\pi}
\right)^n; \nonumber \\
B_{ij}(\alpha_s)&=&\sum^\infty_{n=1}B_{ij}^{(n)}\left(\frac{\alpha_s}{2\pi}
\right)^n;
\label{EQ:ABEXP}  \\
C_{ij}(x,\alpha_s)&=& \delta_{i\bar j}\, \delta (1-x)
+ \sum^\infty_{n=1}C_{ij}^{(n)}\left(\frac{\alpha_s}{2\pi}
\right)^n.
\nonumber
\end{eqnarray}
The reason for the bar over the $j$ in the expression for $C_{ij}(x,\alpha_s)$
is that the flavors of $i$ and $j$ must be the same in the case of $q\bar q$.

We work to first order in the expansions of $A$ and $B$ which corresponds
to summing the first two powers of
$\ln(M^2/q_\perp^2)$ at every order in $\alpha_s$, i.e, the
double-leading logarithm approximation.
The $A_{ij}^{(n)}$'s and $B_{ij}^{(n)}$'s depend implicitly on the choices
of $C_{1,2,3}$.  The simplest forms result from the choices
in Eq.~(\ref{EQ:C123}).  The coefficients $A_{ij}^{(1)}$ and $B_{ij}^{(1)}$
in Eq.~(\ref{EQ:ABEXP}) can be obtained by formally expanding
Eq.~(\ref{EQ:RESUM}) in a series in $\alpha_s$ and then comparing with the
asymptotic perturbative calculation from our previous section.
For initial state gluon radiation, the expressions for $A^{(1)}_{ij}$
in Eq.~(\ref{EQ:ABEXP}) are the ${\cal A}_{ij}^{(1)}$ of Eq.~(\ref{EQ:ACAL}),
and the $B^{(1)}_{ij}$ are the $B^{(1)}_{ij,I}$ in Eq.~(\ref{EQ:BIF}).
For $C$, we make the simplifying choice
$C^{(0)}_{ij}=\delta_{i\bar j}\, \delta (1-x)$ since
we are working in perturbation theory to the first non-trivial
order in $\alpha_s$ for
large $q_\perp$.
(Note that $\sigma_{ij}^{(0)}$ in Eq.~(35) is proportional to $\alpha_s^2$.)
Our neglect of  $C^{(1)}_{ij}$
will only affect
the normalization
at $q_\perp=0$ to ${\cal O}(\alpha_s^3)$ and
the distribution for
$q_\perp \neq 0$ to ${\cal O}(\alpha_s^4)$.
These statements imply that our
calculation includes resummation of the double-leading logarithms to all
orders in $\alpha_s$, but the total integrated cross
section is accurate only to second order in $\alpha_s$.

We comment that the validity of the resummation formalism\cite{cs2}
was demonstrated for the Drell-Yan reaction and for W, Z
production where there is no final state gluon radiation.
We are making the reasonable assumption here that the same formalism is valid
for dealing with the effects of initial state soft and collinear gluon
radiation in the case of heavy quark pair production.
Resummation of soft gluon emission from the final state heavy quarks
has been studied in Ref.~\cite{MW}.

The gluon resummation formula, Eq.~(\ref{EQ:RESUM}),
provides the cross section in the region of small $q_\perp$;
for the high-$q_\perp$ region we use the exact ${\cal O}(\alpha_s^3)$
perturbative calculation.  We will join the results for the low $q_\perp$
and high $q_\perp$ regions using a matching procedure employed previously
\cite{hinch,kauff}.
\begin{eqnarray}
{d\sigma \over dM^2\, dq_\perp^2\, dy}\,({\rm match})&=&
{d\sigma \over dM^2\, dq_\perp^2\, dy}\,({\rm pert})  \nonumber \\
&& +
f(q_\perp/M)\, \left [ {d\sigma \over dM^2\, dq_\perp^2\,dy}\,({\rm resum}) -
{d\sigma \over dM^2\, dq_\perp^2\,dy}\,({\rm asym}) \right].
\label{EQ:FULL}
\end{eqnarray}
The function
\begin{eqnarray}
f(q_\perp/M) = {1 \over 1 +
(3q_\perp /M)^4}
\label{EQ:FMATCH}
\end{eqnarray}
serves to switch smoothly from the matched formula to the perturbative
formula \cite{kauff}.  For details of the matching procedure we refer to
papers by Arnold and Kauffman and subsequent publications.

Before presenting numerical evaluations, we end this section with a few
remarks.  As discussed by Parisi and Petronzio\cite{pp}, the
resummed expression Eq.~(\ref{EQ:WEXP}) is ill-defined when
$b\ge 1/\Lambda_{QCD}$ because confinement sets in and
$\alpha_s$ blows up.  Procedures have been
proposed in the literature\cite{pp,cs2,dsw}
to deal with this difficulty and parameterize
non-perturbative effects.  In this paper we follow the
method used by Collins and Soper\cite{cs2} and by Davies and
collaborators\cite{dsw}.  We replace $W(b)$ in Eq.~(\ref{EQ:WEXP}) by
\begin{eqnarray}
W(b) \rightarrow W(b_*)e^{-S_{np}(b)};
\label{EQ:NPERTW}
\end{eqnarray}
\begin{eqnarray}
b_*=\frac{b}{\sqrt{1+b^2/b_{max}^2}}.
\label{EQ:BSTAR}
\end{eqnarray}
Large values of $b$ are thereby cut-off at some $b_{max}$;
$\exp(-S_{np}(b))$ parameterizes the large-$b$ dependence due to
nonperturbative physics. In principle, $\exp(-S_{np}(b))$ can be
measured, but in practice one can approximate the function with
a simple Gaussian parametrization,
\begin{eqnarray}
S_{np}(b) = b^2[g_1+g_2\ln(b_{max}M/2)].
\label{EQ:SNP}
\end{eqnarray}
According to Davies and collaborators\cite{dsw}
\begin{eqnarray}
g_1=0.15 GeV^2, \; g_2=0.4 GeV^2, \; b_{max}=(2 GeV)^{-1}.
\label{EQ:G12}
\end{eqnarray}
The values of $g_1$ and $g_2$ are obtained by fitting massive lepton-pair
production data at $\sqrt{S}=27$ and $62~$GeV.  There is no strong reason to
believe that the contribution from non-perturbative intrinsic transverse
momentum
should be identical for subprocesses initiated by gluon-gluon scattering,
as in our case, and quark-antiquark scattering, as in massive lepton-pair
production.  When substantial samples of data become available on
$ c\bar c $ and $b\bar b$ production, it should be possible to refine
the choices made here.  Particularly informative in this respect will be
data on the azimuthal angle ($\phi$) dependence.  The extent to which
the quark and antiquark are produced with $\phi$ near $\pi$ is particularly
sensitive to the net transverse momentum
imparted to the quark-antiquark pair\cite{kuebel,mnr}.

\section {Results and Discussion}  \label{3}

In this section we present and discuss some phenomenological applications of
our analysis.  We use HMRS parton
distribution functions\cite{mrs} and the one-loop corrected formula
for the running coupling constant $\alpha_s(\mu)$ with $\Lambda_4=190~$MeV.
The factorization scale and the renormalization scale are chosen
to be the same as the invariant mass of the heavy quark
$Q\bar Q$ pair, i.e. $\mu=M$, unless stated otherwise.

In Fig.~3 we show the lowest order result for the distribution
in invariant mass of a $b\bar b$ pair produced in a
$p\bar p$ collision at the Fermilab collider energy
$\sqrt{S}=1.8~$TeV.
The bottom quark mass is chosen as $m_b=4.75~$GeV.
The distribution peaks at a value of $M$ a few GeV above the
$b\bar b$ pair mass threshold of $2m_b$.
It then decreases quickly as $M$ increases.
This implies that most $b's$ and/or $\bar b's$
are produced at the Fermilab collider
with small momentum in the $b\bar b$ center of mass frame.
Next-to-leading order QCD contributions
change the overall normalization of this curve, but
more important for the sake of our present discussion,
they should not change the shape of the lowest order curve
except very near threshold or far above threshold\cite{real}.

In Fig.~4, the average quantity
$<q_\perp^2(M)>\!{\cal K}$ is plotted
as a function of the pair invariant mass $M$; the factor
$\cal K$ was discussed in Sec.~2.
The quantity $<q_\perp^2(M)>\!{\cal K}$ is
proportional to the square of the average transverse momentum of the
$b\bar b$ pair.
Its value is about 80 GeV$^2$ near threshold and
rises linearly with $M$ in the range shown in the figure.
At the fixed rapidity value $Y=0$, the function has the same shape
and magnitude as in Fig.~4, understandable because the $b\bar b$
pairs are produced centrally.
(An approximately linear rise with $M$ of the
average transverse momentum in the region
$\tau=M^2/S\ll 0.05$ would be expected from simple dimensional arguments.
The growth of the square in Fig.~4 is less rapid than quadratic.)

An important inference may be drawn from Figs.~3 and 4.  We may
estimate from Fig.~3 that the average invariant mass $<M>=15~$GeV.
Glancing at Fig.~4 we notice that near and above
the pair mass threshold $<q_\perp^2(M)>\!\!{\cal K}$ is
larger than 80 GeV$^2$.  Using
${\cal K} \sim 2.4$(see ref. \cite{nde,bnmss}), we deduce that
$b\bar b$ pairs produced at the Fermilab collider are expected to have
an average transverse momentum about $5~GeV$.
As will be discussed below, after integrating over all
values of $M$, we find that
the square of the average transverse momentum $<q_\perp^2>$
is $36.0~GeV^2$ in
the purely perturbative order $\alpha_s^3$ approximation.
Taking the square-root,
we deduce $<q_\perp>_{rms} \sim 6~GeV$.  This value is comparable to, and
slightly larger than, the typical momentum of an individual $b$ or
$\bar b$ in the $b\bar b$ center of mass frame.   Correspondingly,
a significant fraction
(about a quarter to half depending upon how one defines
back-to-back configuration) of $b\bar b$ pairs at the Fermilab collider
are expected to be produced in a configuration that is not back-to-back
in the transverse plane.

The differential cross section $d\sigma/dMdq_\perp$ is presented in
Fig.~5 for three fixed values of mass, $M=15$, $25$ and $50~GeV$.
The dashed curves are from our fixed order $\alpha_s^3$
perturbative calculation.
They are most applicable in the region $q_\perp \simeq {\cal O}(M)$
where there is essentially only one hard scale in the problem.
The fixed order $\alpha_s^3$
results become inapplicable if $q_\perp << {\cal O}(M)$
where, as discussed earlier, the effects of soft gluon contributions must
be incorporated in order to obtain a more reliable result.
The dot-dashed lines show the asymptotic results, Eq.~(29), obtained from the
fixed-order $\alpha_s^3$ results in the limit
$q_\perp \rightarrow 0$.  At small $q_\perp$ the asymptotic
results agree with the perturbative results, as expected.  At larger
$q_\perp$, the asymptotic results manifest unphysical
characteristics that can be traced to the fact that
the functions ${\cal A}_{ij}^{(1)}$ and ${\cal B}^{(1)}_{ij}$
have opposite signs (c.f. Eqns.~(31) and (33)).  Accordingly, we take
the asymptotic results at face value only at small $q_\perp$.
The cross sections obtained from resummation of the effects of initial
soft gluon radiation are shown by the dotted lines.

The resummed results are not expected to follow the
asymptotic results because only the initial-state
soft gluon radiation is included in our resummed formalism.
The resummed and the purely perturbative order $\alpha_s^3$ curves
nearly coincide for $q_\perp$ about 5~GeV and greater, as might be expected
since the mass of the bottom quark is the relevant physical scale at
small $q_\perp$.  Owing to the effects of resummation, the shapes of the
two curves differ significantly for $q_\perp$ less than 5~GeV.

The solid lines in Fig.~5  present our final matched results,
obtained from Eq.~(\ref{EQ:FULL}).  The matched results agree
with the resummed results in the region of
small $q_\perp$ and with the perturbative results at large $q_\perp$.
These three figures demonstrate how the final resummed and matched
results differ from the perturbative results in the small $q_\perp$
region.  They also show that the simple matching procedure seems to
work adequately in our case\cite{catweb}.

To preclude confusion, we
stress that $d\sigma/dMdq_\perp$ is presented in Fig.~5; thus, the
vanishing of the resummed curves as $q_\perp$ goes to zero has a kinematic
origin.  The divergence apparent in Eq.~(2) is not present in the resummed
calculation.

In Fig.~6(a), the perturbative, asymptotic,
and the resummed results from Fig.~5(a) are replotted as
$d\sigma/dM/dq_\perp^2$ versus $q_\perp^2$.  This figure illustrates
the behavior of the cross section at small $q_\perp$ in
a different way, without the phase space factor of $q_\perp$ that
is present in $d\sigma/dM/dq_\perp$ shown in Fig.~5(a).
The same results are plotted again in Fig.~6(b) but with
the switching function Eq.~(41) included as a multiplicative
factor in the asymptotic and resummed
results. A comparison of Fig.~6(a) and 6(b) demonstrates
the effects of the switching function included in the matching formula
Eq.~(40).  At small $q_\perp$, the switching function is close to 1
and does not modify the asymptotic and resummed results.
At large $q_\perp^2\sim 50~$GeV$^2$,
the switching function suppresses the asymptotic and resummed results by
almost a factor of 10.

In Fig.~7 our final results are shown for the distribution in the square of
the transverse momentum of the $b\bar b$ pair, $d\sigma/dq_\perp^2$.
Here we have integrated over $M$.  As remarked at the start of this
section, for consistency, all curves, including
the purely perturbative order $\alpha_s^3$
curve (dashed line), are computed with the one-loop evolved form for
$\alpha_s$

For $q_\perp \neq 0$, the differences between the matched
(solid curve) and fixed-order
$\alpha_s^3$ (dashed) results in Fig.~7 are formally of order
$\alpha_s^4$ and higher(except that our integrated cross section is
valid only to ${\cal O}(\alpha_s^2)$, as explained in Sec.~2).
These differences will affect, among other observables,
the predicted average transverse momentum of the
$b\bar b$ pair.  Computations of the integrals over all $q_\perp$
of the product of $q_\perp^2$ times $d\sigma/dq_\perp^2$ are straightforward
for both the solid and dashed curves in Fig.~7 since this product is finite
and well behaved as $q_\perp$ approaches zero in both cases.  The integral
of $d\sigma/dq_\perp^2$ itself is straightforward for the matched case,
where there is no divergence as $q_\perp$ approaches zero, but is more
involved in the purely $\alpha_s^3$ case.  In the $\alpha_s^3$
case, full account must be taken of virtual diagrams that contribute
at $q_\perp^2$ = 0 \cite{ndet,nde,bkns,bnmss}.  Carrying out the computations,
we obtain $<q_\perp^2>$ = $36.0~$GeV$^2$ in
the purely perturbative order $\alpha_s^3$ approximation and
$<q_\perp^2>$ = $66.7~$GeV$^2$
for our matched case.
We have checked that use of the two-loop evolved form for $\alpha_s$
changes the purely perturbative order $\alpha_s^3$ value of
$<q_\perp^2>$ by less than $1~GeV^2$.

Using the numbers in the paragraph above, we
note that $<q_\perp^2>$ is increased by
30.7~$GeV^2$ as a result of soft-gluon resummation and matching.
Taking square-roots,
we find $<q_\perp>_{rms} \sim 8.2GeV$ in the matched case, to be
compared to $\sim 6~GeV$ in the purely perturbative order $\alpha_s^3$
case.  It may seem remarkable that an additional $\sim 2~GeV$ is
associated with soft-gluon resummation.  It would be useful to be able to
compare this predicted increase with that expected for $<q_\perp^2>$ in
massive lepton pair production (the Drell-Yan process) at Tevatron energies,
at massive lepton pair masses in the vicinity of $10$ to $20~GeV$,
comparable to those for $b\bar b$ pair production.  However, to our
knowledge, no calculations have been published for the resummed $q_\perp$
distribution for the Drell-Yan process at such masses at Tevatron energies.
Calculations at lower energies\cite{cs2,dsw} and/or higher
masses\cite{keith,ak}
may not be a useful guide since the distribution should broaden with energy
at fixed mass.  We remark, however, that the $q_\perp$  distribution should
be significantly broader for $b\bar b$ pair production because the
coefficients in the resummation expression, Eq.~(36), are much larger:
${\cal A}_{gg}^{(1)} = 2 C_A$ vs.
${\cal A}_{q\bar q}^{(1)} = 2 C_F$.

It would be interesting to compare
the solid line in Fig.~7 with data.  We comment
that our result is for production of a pair of $b$ and $\bar b$ quarks,
not a pair of $B$ and $\bar B$ mesons.  To compare
with measurements of the transverse momentum of a pair of
$B$ and $\bar B$ mesons, one must include effects associated with
fragmentation of the $b$ and $\bar b$ quarks, and try to estimate
effects associated with final state gluon radiation.  In a Monte Carlo
simulation of the single $b$ or $\bar b$ inclusive spectrum, Kuebel and
collaborators showed that the effects of final-state and initial state
gluon radiation tended to compensate in some instances, with final-state
radiation tending to soften the spectrum and initial-state radiation
broadening the distribution\cite{kuebel}.  To obtain the $b$ quark inclusive
cross section, the UA1 collaboration\cite{UA1} and the CDF
collaboration\cite{CDFb} use a Monte Carlo procedure to take into account
the fragmentation effects of $b$ quarks into observed $B$ mesons or single
leptons before they compared their data with theoretical predictions.
A similar procedure is required before comparison can be made of
data with our theoretical prediction of the pair transverse momentum
distribution.

In Fig.~8, we show a distribution in the square of the transverse
momentum for production of a pair of charm quarks,
$c\bar c$ pair production in fixed-target proton beam experiment with
$E_{beam}=800~$GeV.  In the case of charm, the average transverse momentum
is so small that non-perturbative physics may dominate in the
region of small $q_\perp$.  We will not dwell here on
applications of the resummation method to charm pair production.  In
another paper, we plan to present further phenomenological applications
for $b\bar b$ pair production.

In this paper, we have focussed on the distribution in transverse momentum
of a pair of heavy quarks.  In carrying out our calculation, we
integrated over (angular) variables in the $Q\bar Q$ rest frame.
Correspondingly, certain limitations must be accepted.  In the context of our
calculation, we are not able to describe the fully differential distribution in
the momenta of the $Q$ and $\bar Q$ separately, notably the azimuthal angle
($\phi$ )dependence in the transverse plane nor correlations in rapidity.
The limitation on the description of rapidity correlations appears
insignificant since these differ little at leading \cite{elb1} and
next-to-leading order \cite{mnr}.  On the other hand, the azimuthal angle
dependence is
sensitive to the net transverse momentum imparted to the $Q\bar Q$ pair
\cite{mnr,kuebel}, and it would be valuable to develop our approach
further in order to examine the influence of
soft-gluon resummation on the $\phi$ distribution.

A calculation of the fully exclusive parton cross section for $Q\bar Q$ pair
production at order $\alpha_s^3$ has been published, along with examples of
distributions
at collider and fixed-target energies \cite{mnr}.  As in the
calculation reported here, those results include the lowest order $\alpha_s^2$
cross section for production of a $Q\bar Q$ pair, the order $\alpha_s^3$
virtual corrections to the lowest order cross section, and the order
$\alpha_s^3$ cross section for production of a $Q\bar Q$ pair along with a
light parton.  That calculation does not include the effects of
soft gluon resummation presented in this paper.  On the other
hand, included in Ref.\cite{mnr} is an exploration of certain other effects
that go beyond the pure $\alpha_s^3$ QCD calculation.  The parton shower Monte
Carlo program HERWIG \cite{herwig} was used to simulate the effects on the
$\phi$ distribution of finite
intrinsic transverse momentum of the initial partons.  Substantial broadening
of the $\phi$ distribution was observed, tantamount to that one would expect if
the incident partons carried an intrinsic transverse momentum of about
1.7 GeV.  While such a large intrinsic contribution was questioned in
Ref.\cite{mnr} as perhaps an unreliable
artifact of the parton shower algorithm, the notable influence of
the added transverse momentum on the $\phi$ distribution underscores the
importance of the type of study carried out in the present paper.  We recall
that our soft gluon resummation introduces substantial additional
$<q_\perp>$.   As a step
in the direction of a full investigation of the influence of soft-gluon
resummation on the $\phi$ distribution, it might be possible to incorporate
the matched distributions shown in Figs.~7 and 8 into a modified order
$\alpha_s^3$ event generator.

In summary, we have studied the distribution in the transverse momentum
of a pair of heavy quarks produced in hadronic reactions.  For
large $q_\perp$, the order $\alpha_s^3$ perturbative result should be
applicable.  In the region of small $q_\perp$, we argued that
resummation techniques developed in the study of the Drell-Yan reaction
should apply for initial state soft gluon radiation.  Use of the resummation
method, plus a matching of results in the small and large $q_\perp$ regions,
permits an improved prediction of the full $q_\perp$ spectrum.
Numerical results for the region of small $q_\perp$ region were presented
for $b\bar b$ pair production at the Fermilab collider and for $c\bar c$ pair
production in fixed target experiments.

\section {Acknowledgments}  \label{5}

We are pleased to acknowledge useful discussions with S. Catani,
J. Collins, R. Kauffman, P. Nason, W. L. van Neerven, J. Smith,
B. Webber, and C-P.~Yuan.  This work was supported in part
by the U.S. Department of Energy, Division of
High Energy Physics, Contract W-31-109-ENG-38.

\pagebreak
%
{}

\pagebreak

\section*{Figure Captions}
\begin{description}
\item[Fig.  1.]
Feynman diagrams at order $\alpha_s^2$.
\item[Fig.  2.]
Examples of order $\alpha_s^3$ Feynman diagrams with
a gluon emitted from an initial quark or gluon line or from
final heavy quark line.
\item[Fig.  3.]
$b\bar b$ pair invariant mass distribution computed from lowest-order
QCD processes.
\item[Fig.  4.]
Average of the square of the transverse momentum of a $b\bar b$ pair,
multiplied by the factor $K$,
as a function of the pair invariant mass $M$.  This curve is
obtained from the purely perturbative order $\alpha_s^3$ calculation.
\item[Fig.  5.]
$b\bar b$ pair transverse momentum distributions for three values of
invariant mass, $M=15, 25$, and $50$ GeV.
We show our fixed order $\alpha_s^3$ perturbative results
as dashed lines, and our initial-state soft gluon resummed results as dotted
lines.  The asymptotic results are represented as dot-dashed lines.
The final matched results are shown as solid lines.
\item[Fig.  6.]
(a) The perturbative, asymptotic,
and resummed results from Fig.~5(a) are plotted as
$d\sigma/dM/dq_\perp^2$ versus $q_\perp^2$.
(b) The same results are plotted again but with
the switching function Eq.~(41) included as a multiplicative factor
in the asymptotic and resummed results.
\item[Fig.  7.]
Distribution in the square of the $b\bar b$ pair transverse momentum.
Our final matched results are shown by the solid line.  For small
$q_\perp^2$ we integrate the leading log resummed result,
Eq.(\ref{EQ:RESUM}), over $M$. The pure ${\cal O}(\alpha^3_s)$
result is represented as the dashed curve, the initial-state soft
gluon resummed result as a dotted
line, and the asymptotic result as a dot-dashed line.
\item[Fig.  8.]
Distribution in the square of the $c\bar c$ pair transverse momentum
for proton-proton collision at $\sqrt{S}=38.7$ $GeV$ (corresponding to
fixed target beam energy 800 GeV).  Curves are labelled as in Fig.~7.
\end{description}

\begin{thebibliography}{}
\bibitem{virtual}
At order $\alpha_s^3$ the
virtual corrections are also proportional to $\delta(q_\perp^2)$ and
therefore do not contribute to the cross section
at any finite $q_\perp^2\ne 0$.
\bibitem{LSN}
A more detailed discussion on this
point can be found e.g. below Eq.~(2.9) in Ref.~\cite{LSN1}
and in ref.~\cite{ddt}.
\bibitem{real}
We have checked that the contribution to $d\sigma/dM^2$ from the
processes~(\ref{EQ:ALFSP}), after a $q_\perp > 5~$GeV cut is applied, agrees
with the shape of the lowest order result in the $M^2$ range shown in
the figure.
\end{thebibliography}
\end{document}